\documentclass[preprint,12pt]{aastex}
\slugcomment{Submitted to ApJ August 16, 2007; Accepted Jan 17, 2008}

\newlength{\figsize}
\def\eq#1{\begin{equation} {#1} \end{equation}}
\def\sub#1{_{\rm #1}}
\def\E#1{\hbox{$10^{#1}$}}
\def\about  {\hbox{$\sim$}}
\def\x      {\hbox{$\times$}}
\def\ga     {\hbox{$\gtrsim$}}
\def\la     {\hbox{$\lesssim$}}
\def\mic    {\hbox{$\mu$m}}

\def\DV     {\hbox{$\Delta V$}}

\def\kms    {\hbox{km\,s$^{-1}$}}
\def\cc     {\hbox{cm$^{-3}$}}
\def\cmsq   {\hbox{cm$^{-2}$}}

\def\LOH     {\hbox{$\cal L\sub{OH}$}}
\def\LHCN     {\hbox{$\cal L\sub{HCN}$}}
\def\LHCO+     {\hbox{$\cal L\sub{HCO^+}$}}
\def\LIR    {\hbox{$\cal L\sub{IR}$}}

\def\PI(#1,#2){\hbox{$^2\Pi_{#1}(J = #2)$}}
\begin{document}

\title{The Effect of 53 \mic\ IR Radiation on 18 cm OH Megamaser Emission}

\author{Philip Lockett\altaffilmark{1} and Moshe Elitzur\altaffilmark{2}}

\altaffiltext{1}{Centre College, 600 West Walnut Street, Danville, KY 40422;
lockett@centre.edu}

\altaffiltext{3}{Physics and Astronomy, University of Kentucky, Lexington, KY
40506-0055; moshe@pa.uky.edu}

\begin{abstract}

OH megamasers (OHMs) emit primarily in the main lines at 1667 and 1665 MHz, 
and differ from their Galactic counterparts due to their immense 
luminosities, large linewidths and 1667/1665 MHz flux ratios, which are 
always greater than one. We find that these maser properties result from 
strong 53 \mic\ radiative pumping combined with line overlap effects caused 
by turbulent linewidths $\sim$ 20 \kms; pumping calculations that do not 
include line overlap are unreliable. A minimum dust temperature of \about\ 
45 K is needed for inversion, and maximum maser efficiency occurs for dust 
temperatures \about\ 80 -- 140 K. We find that warmer dust can support 
inversion at lower IR luminosities, in agreement with observations. Our 
results are in good agreement with a clumpy model of OHMs, with clouds 
sizes \la\ 1 pc and OH column densities \about\ 5\x\E{16}\cmsq, that is 
able to explain both the diffuse and compact emission observed for OHMs. We 
suggest that {\em all} OH main line masers may be pumped by far-IR 
radiation, with the major differences between OHMs and Galactic OH masers 
caused by differences in linewidth produced by line overlap. Small Galactic 
maser linewidths tend to produce stronger 1665 MHz emission. The large OHM 
linewidths lead to inverted ground state transitions having approximately 
the same excitation temperature, producing 1667/1665 MHz flux ratios 
greater than one and weak satellite line emission. Finally, the small 
observed ratio of pumping radiation to dense molecular gas, as traced by 
HCN and HCO$^+$, is a possible reason for the lack of OH megamaser emission 
in NGC 6240.

\end{abstract}


\section{Introduction}

OH megamasers are extremely luminous extragalactic OH maser sources that have
isotropic luminosities a million or more times greater than their Galactic
counterparts. OHMs are found only in the nuclear regions of luminous (LIRGs)
and ultraluminous (ULIRGs) infrared galaxies, where intense star formation is
occurring. These masers emit primarily in the main lines at 1667 and 1665 MHz,
although weaker emission in the satellite lines at 1612 and 1720 MHz has also
been detected \citep{baan87}. In addition to their immense luminosities, OHMs
differ from their Galactic counterparts in their very large linewidths and
their 1667/1665 MHz flux ratios. Galactic OH mainline masers in star forming
regions tend to have linewidths \la\ 1 \kms\ and 1667/1665 MHz flux ratios less
than one. In contrast, OHMs have overall linewidths \ga\ 100 \kms\ and their
1667/1665 MHz flux ratios are always greater than one. Although this ratio
varies, the typical ratio is \about\ 5 \citep{lonsdale02}. \citet{baan89} found
a clear separation between OH emitters and absorbers based on the IR properties
of the host galaxies. The OHMs have larger IR luminosity (\LIR) and tend to be
warmer. OHMs may occur for lower \LIR\ as long as the dust is warm enough. The
Arecibo OHM survey conducted by \citet{darling02} targeted LIRGs and increased
the number of OHMs to \about\ 100. That survey substantiated the conclusions of
\citet{baan89} and showed that the fraction of OHMs in LIRGs is an increasing
function of far-IR (FIR) luminosity and color, with most of the detections
found in warm ULIRGs. These observations strongly suggest that the masers are
pumped by the intense FIR radiation in the OH pump lines at 53 \mic\ and 35
\mic. Support for FIR pumping is provided by observations using the Infrared
Space Observatory (ISO). \citet{skinner} observed the 35 \mic\ pumping
transition in Arp 220 and concluded that absorption in this line alone could
power its maser emission. Significantly, OH and water megamasers have not been
observed in the same galaxy \citep{lo05} even though both are often found in
the same Galactic star forming regions.

The standard OHM model was introduced by \citet{baan85}, who proposed that the
maser emission is produced by low gain unsaturated amplification of background
radio continuum. This proposal was confirmed by the comprehensive study of OHMs
performed by \citet{henkelwilson90}. They found good agreement with
observations if the maser transitions have approximately equal excitation
temperatures, giving a 1667/1665 MHz optical depth ratio of \about \ 1.8.
However, subsequent VLBI observations revealed compact maser emission on parsec
scales with amplification factors \ga \ 800 and very large 1667/1665 MHz line
ratios \citep{lonsdale98}. A surprise of these observations was that linewidths
remained large (tens of \kms) even on the smallest observed angular scales
\citep{lonsdale02}. The VLBI observations led to the suggestion that there are
two different modes of maser operation --- low gain, unsaturated amplification
responsible for the diffuse maser emission and high gain, saturated emission
producing the compact sources. It was also suggested that the two classes may
have different pumping mechanisms, with the diffuse emission pumped by IR
radiation and the compact masers by a combination of collisions and radiation
\citep{lonsdale02}. However, an extension of the standard model is able to
explain both the diffuse and compact emission from IIIZw35 by assuming a clumpy
maser medium \citep{parra05}. Each cloud generates low gain, unsaturated
emission, and strong compact emission occurs when the line of sight intersects
more than one cloud. A similar model has been used to explain the megamaser
emission from IRAS 17208-0014 \citep{momjian}. The clumpy maser model also
explains the observation that compact masers are always found embedded in the
diffuse emission and do not occur in isolation \citep{lonsdale02}.

An observational difficulty is that the nearest OHMs are located at distances
of \about \ 100 Mpc and thus cannot be probed nearly as accurately as Galactic
masers. The study of OHMs thus relies heavily on analyzing their global
properties. It is important to first explain the global properties of OHMs and
then construct more comprehensive models for those sources which have been
observed in more detail using VLBI techniques. Previous OHM models
\citep{henkelwilson90,randell} were performed before the discovery of the
compact maser sources. Although the clumpy maser model seems to provide a
phenomenological explanation to the compact and diffuse maser emissions, it is
not yet backed by pumping calculations. The time is right to develop a pumping
model that can explain both the compact and diffuse emission using physical
conditions that are consistent with those expected in the maser regions. This
is the aim of the present paper. In section 2 we summarize what is known
concerning the physical conditions existing in the maser region. In section 3
we explain our model and present the results of our calculations. Section 4
contains a summary and discussion.

\section{Physical Conditions in the Maser Region}

Calculating maser intensities requires knowledge of the physical conditions
existing in the maser region. The important parameters are the FIR radiation
intensity, OH column density, gas density and temperature, and linewidth. The
physical conditions in the nuclear regions of OHMs are significantly different
from those found in normal galaxies due to the extreme amount of massive star
formation. Although hidden from view at visible wavelengths, \citet{lonsdale06}
have observed numerous radio point sources in the nuclear region of Arp 220 and
conclude that they are radio supernovae. They deduce a supernova rate of \about
\ 4 per year, yielding a per-volume rate that is orders of magnitude larger
than Galactic. These radio sources have recently been observed at other radio
wavelengths and the SN nature of the sources has been strengthened
\citep{parra07}. That study also found that an unusually large fraction of
these supernovae are highly luminous, suggesting that stellar formation and
evolution may be much different in these regions compared to normal galaxies.
The large amount of massive star formation results in the intense FIR radiation
and turbulence present in the region of maser production.

\subsection{FIR Radiation Intensity}

The intense FIR radiation field present in the nuclei of LIRGs and ULIRGs is
the dominant pump for the OHMs. Estimates of the effective dust temperature and
opacity have been calculated in numerous studies. The IRAS derived color
temperatures of 74 OHMs range from 40 -- 90 K with a median of 60 K
\citep{klockner}. \citet {yuncarilli} find best fit dust temperatures in the
range 59 -- 72 K for the OHMs Arp 220, Mrk 273 and Mrk 231. These are global
estimates and may be influenced by cooler dust lying outside the maser region.
\citet{soiferarp} performed high spatial resolution observations of the nuclear
region of Arp 220 from 3 to 25 \mic\ and conclude that the region is
complicated with a range of source temperatures and sizes. They find that
emission at wavelengths longer than 20 \mic\ corresponds to a dust temperature
of \about \ 85 K. This result is consistent with a detailed theoretical
analysis of radiation pressure-supported starbursts, which finds the effective
dust temperature to be $\sim$ 90 K for the optically thick starburst disks of
ULIRGs \citep{thompson}. A possible upper limit to the dust temperature of
\about \ 100 K may be provided by the lack of observation of the 6.7 GHz
methanol maser emission in OHMs \citep{darlingmethanol}. This methanol
transition is often found to be a strong maser in association with OH masers in
Galactic star forming regions. Modeling of the 6.7 GHz maser finds that it is
radiatively pumped and requires dust temperatures \ga \ 100 K \citep{cragg}.

The dust opacity in OHMs is not constrained as tightly by observations, and the
estimated opacity depends on the wavelengths being examined.
\citet{lisenfeld00} discuss a number of arguments that suggest visual opacities
of order 1000. However, other studies estimate much smaller visual opacities in
the range 20-200 \citep{soiferarp, hinz}. It should be noted that the values of
dust temperature and opacity that best fit the observed spectral energy
distributions are not independent and there is considerable degeneracy between
the two parameters \citep{blain}.

\subsection{Gas Temperature and Molecular Density}

Gas temperature and density are needed to calculate collision rates. Although
OHM pumping is dominated by radiation, collisions need to be included in the
calculation of the OH level populations. It has also been suggested that the
compact masers may be pumped by a combination of collisions and radiation
\citep{lonsdale02}. The temperature and density can be estimated using high
density tracer molecules such as HCN, CS and HCO$^+$. \citet{cernicharo}
observed transitions in HCN and HNC in Arp 220 and their analysis predicts
densities in the range \E5--\E6 \cc\ and gas temperature \about\ 40 K.
\citet{greve} observed Arp 220 using numerous molecular lines including those
of HCN, CS and HCO$^+$. They deduce densities in the range \E4--\E6 \cc\ and
gas temperatures ranging from 50 K to 70 K. Further constraints on these
parameters are provided by the lack of detection of the 22 GHz water maser
combined with the detection of the 183 GHz water maser in Arp 220
\citep{cernicharo}. The 22 GHz maser is collisionally pumped and requires
relatively high densities and temperatures for its production. The 183 GHz
maser is produced at much lower densities and temperature. \citet{cernicharo}
perform calculations that show that strong 183 GHz emission without 22 GHz
emission requires densities \la\ \E6 \cc\ and gas temperatures \la\ 100 K.

\subsection{Linewidth}
\label{sec:width}

Linewidth usually affects the level populations only as part of the combination
$n_{\rm OH} R/ \Delta V$, the column density per velocity interval; neither
$n_{\rm OH}$, $R$ nor $\Delta V$ enter separately. However, overlap between FIR
rotational lines of OH introduces the linewidth as an independent parameter
with additional, separate effects on the level populations. Line overlap
affects photon trapping since a photon emitted in one transition may be
absorbed in a different one. In addition, line overlap can significantly affect
the absorption of the pumping radiation in the overlapping transitions. The
lowest 24 hyperfine levels of OH are shown in Figure \ref{fig:levels}. The
level structure is more complicated than is found for the typical diatomic
molecule. There are two rotational ladders, and each rotational level is split
into 4 sublevels due to the effects of lambda doubling and hyperfine splitting.
The hyperfine separations differ in the 2 halves of the lambda doublets leading
to line separations as small as .1 \kms\ between FIR rotational lines. Figure
\ref{fig:levels} shows that the asymmetry is especially large for the
transitions involving the $^2 \Pi _{1/2}$ ladder.

An important feature of OHMs is their very large linewidth. Single dish
observations of OHMs reveal linewidths of hundreds of \kms. Much of this
linewidth is due to large scale motions of individual maser clouds. However, VLBI imaging
finds that linewidths are still tens of \kms \ even for parsec size regions
\citep{lonsdale02}. These observations indicate the presence of a large amount
of supersonic turbulence and are in agreement with the theoretical predictions
of \citet{thompson}, who estimate turbulent linewidths of $\sim $ 50 \kms.
\citet{downes} performed a detailed analysis of the central region of ULIRGs
using CO interferometry and found that the contribution of local turbulence to
the overall linewidth is $\gtrsim$ 30 \kms. Detailed modeling of the megamasers
in IIIZw35 by \citet{parra05} found best agreement with observations using a
cloud internal velocity dispersion of 20 \kms. These large linewidths lead to a
significant amount of overlap in the OH FIR transitions.

\section{Modeling and Results}

Although all four ground state transitions have been observed as masers in
OHMs, the satellite lines at 1612 MHz and 1720 MHz are much weaker than the
main lines at 1667 MHz and 1665 MHz. Satellite line maser pumping follows
naturally from the OH level structure, but main line pumping requires an
asymmetry in the excitation of the two halves of the lambda doublets
\citep{elitzur}. Possible asymmetric pumps include radiative pumping by warm
dust or overlap between the OH FIR rotational lines. In the Appendix we discuss
FIR pumping in detail and show that line overlap completely dominates all other
suggested FIR pumping mechanisms of main line masers.

We solve for the populations of the 24 levels shown in Figure \ref{fig:levels}.
These include all the levels with direct radiative connection to the ground
state, where the masers are produced. We have checked that increasing the
number of levels makes no difference on the outcome. The statistical
equilibrium equations are set in the escape probability approach using the
\cite{capriotti} escape probability. This is equivalent to a slab approximated
with a constant source function, and should provide reasonable estimates of the
overall emission from the region. Further discussion of this technique and
comparison with exact radiative transfer solutions can be found in
\cite{elitzur06}. As noted in \S\ref{sec:width}, the OH energy level structure
necessitates inclusion of the effects of line overlap. We use the method of
\citet{lockettelitzur89} to treat the local overlap of the FIR rotational
lines. This procedure is based on the slab escape probability method and
accounts for overlap effects by numerically integrating the amount of
absorption in the overlapping lines assuming Gaussian line profiles. This
method is similar to that of \citet{guilloteau}, except they assume a spherical
geometry as opposed to a slab.

We envision the maser source embedded in a dusty environment that provides the
pumping IR radiation. Assuming the dust temperature $T_d$ to be approximately
uniform across the maser cloud, the local radiation field can be described with
\eq{ I_\nu = B_\nu (T_d)\left(1-e^{-\tau_{\nu}}\right) } Here $B_\nu (T_d)$ is
the Planck function and $\tau_{\nu}$ is the frequency dependent dust opacity
across the region where its temperature can be described by the single value
$T_d$. Note that $\tau_{\nu}$ is the dust optical depth over the entire dusty
region and can be larger than its value across the maser cloud. The spectral
shape of $\tau_\nu$ is found using the cross sections of \cite{drainelee84}.

Although radiation is the dominant pumping mechanism, collisions need to be
included in the calculations and we utilize the most recent collision rates
from \cite{offeretal94}. These rates are considerably different for collisions
with ortho and para $H_2$, and the pumping could depend sensitively on the
ortho-para ratio \citep{lockettetal99}. This ratio is uncertain in molecular
clouds. $H_2$ is believed to form with the ortho-para ratio \about\  3 and the
ratio approaches thermal equilibrium at a rate depending on the chemical
history of the cloud \citep{flower}. We performed calculations ranging from
pure ortho to pure para and the results did not vary significantly. Our nominal
choice for the ortho-para ratio is one, which is the equilibrium value for T
\about \ 77 K.

We explored a wide range of the physical parameters and found a large volume of
phase space where the main lines are inverted. Table \ref{table:model} lists
the values of the parameters we consider ``standard". We now present our
results, varying the model parameters one at a time from these nominal values.

%%%%%%% Table - model parameters
\clearpage
\begin{table}[ht]
\centering

\begin{tabular}{ll}

\tableline \tableline

Dust temperature, $T_d$            & 60 K \\
Dust optical depth at visual        & 100 \\
Linewidth (FWHM), $\Delta V$       & 20 \kms \\
OH column density, $n_{\rm OH}R$   & 6\x\E{16} \cmsq \\
Gas temperature                    & 50 K \\
Gas density                        & \E4\ \cc \\

\tableline

\end{tabular}

\caption{``Standard" model parameters chosen to be representative of individual 
clouds in the clumpy OHM model of \citep{parra05}. Figures show the effects of 
varying individual parameters from these nominal values, as indicated in the
corresponding captions.}

\label{table:model}
\end{table}
\clearpage
%%%%%%%%%%%%%%%%%%%%%%%%%%%%%%%%%%%%%%%%%%%%%%%%%%%%%%%%%%%%%%%%%%%

\subsection{Effects of dust temperature and opacity}
 \label{sec:dust}

Figure \ref{fig:tau contours} plots the contours of 1667 MHz negative optical
depth versus dust temperature and opacity. A minimum dust temperature of about
 45 K is needed for maser production, and maximum maser inversion occurs for
dust temperatures between 80 -- 140 K for the expected range of dust opacity.
When the dust opacity becomes very large, the dust radiation approaches that of
a blackbody and the maser begins to thermalize. Beyond a visual opacity of
about 300 there is little dependence of maser strength on dust temperature.
There is also a minimum FIR pumping flux needed for maser production, that is a
function of dust temperature. The pumping flux  in equation 1 is determined by
the product of two factors: the Planck function, which depends on dust
temperature, and the self-absorption factor, which depends on the dust opacity
at the pumping frequency. The FIR flux is increased by increasing either the
dust temperature or opacity. Figure 3 shows the minimum 53 \mic\ flux needed
for inversion as a function of dust temperature. A larger flux is needed for a
lower dust temperature, indicating that the FIR spectral shape is important for
maser pumping. Thus a smaller \LIR \ can support inversion, as long as the
temperature is warm enough. This agrees with the observations of \citet{baan89}
and \citet{darling02}. Our analysis also shows that a minimum dust temperature
is needed for inversion. Below 50 K, the Planck function at 53 \mic\ decreases
very rapidly as dust temperature decreases. The flux can be increased by
increasing dust opacity, but it cannot exceed blackbody emission and the needed
flux can not be produced for dust temperatures below about 45 K.

The pumping is primarily by the 53 \mic\ lines, with the 35 \mic\ transitions
in a secondary role. Increasing the number of OH energy levels in the model
calculations, main-line inversion was not produced until the $^2\Pi_{1/2}(J =
3/2)$ levels, at 53 \mic, were included. The 53 \mic\ lines dominate the
pumping, but the maser strength increased further when the four levels of
$^2\Pi_{1/2}(J = 5/2)$, at 35 \mic, were added. We find the opacity of the 53
\mic\ lines to be about 5 times those of the 35 \mic\ lines. This agrees
with the study of \citet{he} who compared the line strengths in Arp 220 and
found the amount of absorption in the 53 \mic\ line to be about 6 times that
of the 35 \mic\ line. Results were hardly affected by inclusion of the
$^2\Pi_{3/2}(J = 7/2)$ levels, as expected since these levels are radiatively
disconnected from the ground state.

An important observed correlation is the increase of \LOH\ with \LIR. Early OHM
observations revealed an approximately quadratic relation between the two
\citep{baan89}, but subsequent studies using a much larger sample and taking
into account Malmquist bias established the flatter dependence $L_{OH}
\varpropto L_{IR}^{1.2}$ \citep{darling02}. A linear relation is expected
because low gain amplification of background radio continuum implies that \LOH\
is proportional to the radio luminosity, which in turn is proportional to \LIR\
over the relevant luminosity range \citep{darling02}. This linear
proportionality will not be altered by the expected variations among different
sources in OH covering factor of the radio continuum, which will only induce a
scatter around it. Since the observed correlation is steeper than linear,
\cite{baan89} conjectured that maser optical depth $\tau_{\rm maser}$ also
varies linearly with the pump luminosity \LIR. However, our calculations show
that, depending on the different parameters, $\tau_{\rm maser}$ in fact can
either increase or decrease with the IR luminosity. Indeed, in their extensive
study of OHMs, \citet{henkelwilson90} did not find a correlation between \LIR \
and $\tau_{\rm maser}$. A possible explanation for the increase of \LOH\ in
excess of linear proportion to \LIR\ may be provided by the study of HCN
emission from LIRGs by \citet{gao}. They find that \LIR\ $\propto$ \LHCN\ and
conclude that the amount of dense ($n_{H_2} $ \ga\ \E4 \cc ) molecular gas,
which is traced by the HCN, is linearly proportional to \LIR. It is thus
reasonable to assume that, together with the increase of dense molecular gas,
the number of OH maser clouds is also increasing with \LIR.

\citet{lo05} pointed out that, based on its FIR color and \LIR\ the ULIRG NGC 
6240 should be an OHM \citep[see also figure 2 of][]{baan89}. A possible reason 
why it is not is its low ratio of \LIR\ to dense molecular gas. This gas is 
traced by HCN and HCO$^+$ molecules, and Table 1 in \citet{carpio} shows that 
the ratios \LIR /\LHCN \ and \LIR /\LHCO+ \ are both less than half in NGC 6240 
than in every OHM, seven in all, included in their sample. Thus the FIR pump 
rate per molecule in NGC 6240 is much lower than in the typical OHM. Therefore, 
a minimum ratio of pumping radiation to dense molecular gas may also be a 
prerequisite for producing OHM. \citet{darling07} has recently conducted a 
detailed study that shows a clear separation between OHMs and OH absorbers based 
on the amount of dense molecular gas.

\subsection{Effects of collisions}
 \label{sec:collisions}

We found no conditions where collisions alone produce the observed maser
emission. When combined with a strong FIR radiation field, collisions tend to
weaken and then thermalize the masers. Increasing the collision rate by
increasing either gas temperature or density tends to reduce the maser
strength. An increase in the collision rate also increases the dust temperature
required for inversion --- increasing the density from \E4 to \E5 \cc\ raises
the minimal dust temperature from about 45 K to 50 K. These results argue
against collisional pumping as a source of the compact masers. Another argument
against collisional pumping is the absence of the 22 GHz water maser combined
with the presence of the 183 GHz water maser in Arp 220. We calculated the
level populations for ortho and para water for conditions representative of the
OHM region, and our results are consistent with those of \citet{cernicharo}.
The presence of the 183 GHz maser combined with the absence of the 22 GHz maser
implies densities \la\ \E6 \cc\ and temperature \la\ 100 K. These conditions
are not conducive to collisional pumping of the OHM.

\subsection{Effects of Line Overlap}
\label{sec:overlap}

We find that the effects of line overlap dominate the main line inversion even
at linewidth as small as 0.5 \kms. In the Appendix we discuss FIR pumping in
detail and show that line overlap completely dominates all other suggested FIR
pumping mechanisms of main line masers. Calculations that do not include
overlap effects cannot produce reliable results for FIR pumping.

Figure \ref{fig:linewidth} shows the effects of line overlap on the 1667 MHz
maser strength and the 1667/1665 MHz optical depth ratio. Because optical depth
is determined by the column density per velocity interval, $n_{\rm OH}R/\Delta
V$ remains a scaling variable, as is evident from the figure's right panels;
they show that the pumping results display similar behavior for the different
linewidths when plotted in terms of this variable. Line overlap effects add
$\Delta V$ as another independent parameter of the pump scheme, with striking
effects on the inversion: the model results for the ratio $R =
\tau(1667)/\tau(1665)$ split into two groups, with a sharp transition between
the two. Starting from $\Delta V$ = 2 \kms, a decrease of 1 \kms\ has little
effect on $R$ while an increase of 1 \kms\ increases $R$ significantly,
especially at the lower end of column densities. Linewidths \ga\ 10 \kms\ have
$R \simeq 1.8$ independent of \DV. All ground state hyperfine transitions are
then inverted with approximately the same excitation temperature, an
equality that was also noted by \citet{henkelwilson90}. In contrast, linewidths
\la\ 2 \kms\ produce $R\  \la\ 1$, again displaying behavior that is nearly
independent of \DV. Further discussion of this behavior is provided in the
Appendix.

Our results can explain why the flux ratios observed for OHMs differ radically
from those found for their Galactic counterparts. Galactic interstellar masers
tend to have small linewidths and the 1665 MHz line is usually the strongest,
with the strength of the 1667 MHz line often comparable to that of the
satellite lines \citep{gaume}. However, there are exceptions where the 1667 MHz
line or one of the satellite lines is the strongest. Our results show that this
is the expected behavior for \DV\ \la\ 2 \kms. In contrast, in OHMs the 1667
MHz line is always the strongest and the satellite lines are much weaker by
factors \ga \ 10, the expected behavior when \DV\ \ga\ 10 \kms. Significantly,
main line masers in OH/IR stars tend to have larger linewidths than Galactic
interstellar masers and their 1667/1665 MHz flux ratios tend to be larger than
one.

We find that all ground state transitions are inverted with approximately equal 
excitation temperatures when \DV\ \ga\ 10 \kms. This is consistent with the OHMs 
global fluxes determined from observations \citep{henkelwilson90}. The optical 
depth of the 1667 MHz line is then \about \ 1.8 times that of the 1665 MHz line 
and \about\ 9 times that of the satellite lines and the fluxes for the OH maser 
amplified background continuum become functions of optical depth alone. Figure 
\ref{fig:fluxes} displays the amplified 1665, 1667 and 1720 MHz line fluxes. 
Because of its stronger amplification, at a given instrumental sensitivity the 
1667 line will be detected over the largest portion of a source. The 1667/1665 
and 1667/1720 flux ratios increase as the 1667 MHz maser gets stronger. Small 
optical depths give a 1667/1665 MHz flux ratio of \about\ 2 and the typically 
observed ratio of 5 implies the magnitude of $\tau(1667)$ is \about\ 3.5. 
Because the 1720 MHz is so weak compared to the 1667 MHz line, its detection is 
limited. Figure \ref{fig:color-color} plots the track of 1667/1720 MHz flux ratio 
versus the 1667/1665 MHz flux ratio as $\tau(1667)$ is varied. Also shown are the 
data for four relatively nearby OHMs, taken from \citet{baanetal89}, \citet{baanetal92} 
and \citet{martinetal89}. The agreement is surprisingly good considering the 
large uncertainties in the data.

\subsection{Clumpy OHM}
 \label{sec:clumps}

The global properties of OHMs are consistent with the standard model of low
gain amplification of background radiation with the maser transitions having
approximately equal excitation temperatures. Compact, bright maser emission is
explained by the clumpy nature of the amplifying medium. Each maser cloud
produces low gain, unsaturated emission. Compact emission would be observed
when the line of sight intersects a number of maser clouds. This model has been
used to give very good agreement with the maser emission from IIIZw35.
\citet{pihl} observed IIIZw35 using intermediate and high resolution, and
detected both diffuse and compact emission. They proposed a model based on a
clumpy maser medium located in a rotating ring. This model was extensively
improved by \citet{parra05}, who used Monte Carlo simulations to account for
the overlap of individual maser clouds. The clumpy maser model gives results
that are in excellent agreement with the observed spectra and with
interferometric observations on all angular scales. The compact maser emission
arises at the tangents to the ring where cloud overlap is greatest in both
position and velocity. The model of IIIZw35 finds best agreement with
observations using maser clouds having a size \la \ .7 pc, magnitude of $\tau(1667$) 
\about\ 1.5 and velocity dispersion of \about \ 20 \kms. A similar,
but less detailed, model has since been developed to explain the compact and
diffuse emission from IRAS17208-0014 \citep{momjian}.

Our ``standard" parameters in Table \ref{table:model} were selected to match
the IIIZw35 model results of \citet{parra05}. Figure \ref{fig:clump} is an
expanded view of the relevant portion in figure \ref{fig:linewidth} and shows that
the magnitude of $\tau(1667$) is about 1.5 at an OH column density of \about\
5\x\E{16} \cmsq. With the model density of \E{4} \cc, this implies an OH
abundance of \about\ 2\x\E{-6} for the cloud size of 0.7 pc. Calculated OH
column densities are typically in the range \E{15} to \E{16} \cmsq\ for both
PDRs \citep{hartquiststernberg} and C-shocks \citep{wardle}. Larger OH column
densities may be obtained through other mechanisms such as the
photodissociation of water produced from the evaporation from grain mantles
\citep{hartquistetal}. However, even that mechanism has an upper limit of
\about \E{17} \cmsq. Figure \ref{fig:linewidth} shows that very large 1667 MHz
optical depths can be produced, but only if \DV\ is \la\ 20 \kms\ or
the OH column exceeds \about\E{17} \cmsq. Neither condition is likely to be met
in OHMs. Maser amplification by single clouds is weak because linewidths are
large (\ga\ 20 \kms) and OH columns are limited to \la\ \E{17} \cmsq.

The clumpy maser model can also be used to explain the lack of compact maser
emission from Mrk 231. This is the only OHM examined with VLBI not to exhibit
compact emission \citep{lonsdale03}. Furthermore, the 1667/1665 MHz flux ratio
has been found to be \about\  2 on angular scales ranging over 3 orders of
magnitude \citep{richards, klockneretal}. These observations are consistent
with the main lines having similar excitation temperatures while undergoing low
gain amplification. Speculative explanations for the lack of compact emission
have been proposed \citep{lonsdale03}, but a simpler explanation is based on
the geometry of a rotating clumpy maser disk. \citet{richards} and
\citet{klockneretal} each propose that the masers are produced in a rotating
disk that is more inclined to the line of sight than is found for OHMs such as
IIIZw35. The length along the line of sight through the disk is thus smaller
than if the disk were observed edge on, resulting in little cloud overlap and
no compact emission. \citet{richards} estimate maser cloud sizes \about 1 pc
with the magnitude of $\tau(1667$) about 1. These estimates are comparable to
the values suggested by \citet{parra05} in their detailed analysis of IIIZw35.

\section{Summary and Discussion}

OHMs are characterized by their extreme luminosity, 1667/1665 MHz flux ratio
greater than one, and large linewidths. Our detailed pumping analysis of OHMs
shows that the above properties are a natural outcome of an intense FIR pump
combined with line overlap effects due to large turbulent linewidths. The 53
\mic\ lines provide the primary pump, and a minimum dust temperature of \about \
45 K is needed for main line maser production. In agreement with the
observations of \citet{baan89} and \citet{darling02}, we find that a smaller
\LIR \ can support inversion, as long as the temperature is warm enough. We find
no conditions where collisions can produce the observed maser emission, and when
combined with a strong FIR radiation field, collisions tend to weaken and then
thermalize the masers. These results are consistent with the detection of the 183
GHz water maser combined with the lack of detection of the 22 GHz water maser
in Arp 220 \citep{cernicharo}. The 22 GHz maser is collisionally pumped and
requires relatively high densities and temperatures for its production, while the
183 GHz maser is produced at much lower densities and temperature. The large OHM
linewidths produce significant overlap among the the 53 \mic\ pump lines
resulting in nearly equal negative excitation temperatures for all ground state
transitions, in agreement with observations \citep[figure \ref{fig:color-color};
see also][]{henkelwilson90}. The 1667/1665 MHz flux ratios are thus greater than
one and only weak satellite line emission is produced. Our pumping results are
in agreement with those required by clumpy maser models that explain both the
diffuse and compact maser emission from OHMs such as IIIZw35. Finally, the small
observed ratios of \LIR /\LHCN \ and \LIR /\LHCO+ \ relative to those found in
OHMs is a possible reason for the lack of OH megamaser emission in the ULIRG
NGC6240. A minimum ratio of pumping radiation to dense molecular gas may be a
prerequisite for becoming a OHM.

Excited state transitions at 5 cm and 6 cm have also been observed in OHMs.
However, these lines have only been detected in absorption. \citet{henkel87}
observed absorption in the three 6 cm $^2\Pi_{1/2}(J = 1/2)$ transitions in
five OHMs, and absorption has been detected also in one 5 cm $^2\Pi_{3/2}(J =
5/2)$ transition in Arp 220 \citep{henkel86}. These observations are surprising
since masing in the 5 cm and 6 cm lines has often been observed in conjunction
with Galactic ground state OH maser sources \citep{cesaroni}. Our pumping
models of ground state maser emission in OHMs usually predict absorption in the
5 cm lines. However, the 6 cm lines are weakly inverted with opacities a
hundred times smaller than the 1667 MHz line. Our calculations find that
absorption in the 6 cm lines requires smaller dust temperatures than exist in
the maser regions. We expect the absorption to occur outside the maser regions
where the dust temperature is lower. Understanding the absorption at 5 cm and 6
cm is an important problem that we plan to treat in a future paper.

\subsection{OH main-line maser pump}

Our results suggest that {\em all} OH main-line masers could be pumped by the
same mechanism: far-IR radiation. The major differences between Galactic star
forming regions on one hand and evolved stars and OHMs on the other can be
attributed to differences in line overlap effects. As shown in
\S\ref{sec:overlap}, linewidth is the most important factor in determining the
1667/1665 MHz flux ratio and thus is the dominant factor causing the extreme
differences between the flux ratios of the two classes. In Galactic star
forming regions the observed linewidths are \la \ 1 \kms\ and the 1665 MHz
maser is usually strongest, with the 1667 MHz line often comparable in strength
to the satellite lines. In OHMs the linewidths are large (\ga \ 10 \kms) and
the 1667 MHz is always stronger than the 1665 MHz line, with the satellite
lines much weaker. Our results show that such large linewidths produce
approximately equal negative excitation temperatures for the four ground state
transitions, resulting in the 1667 MHz line being stronger than the 1665 MHz
line and much weaker satellite lines. Smaller linewidths tend to produce a
stronger inversion in the 1665 MHz line. In evolved stars, the main-line maser
velocities are larger than those found in Galactic star forming regions and
they also tend to have 1667/1665 MHz flux ratios \ga\ 1. Thus the linewidth
could be the main reason why OH main-line masers display differences in spite
of sharing a common pump mechanism.

\acknowledgements
Partial support by NSF grant AST-0507421 (M.E.) is gratefully acknowledged.

\appendix
\section{Main-line Pumping by FIR Radiation }

%\section{Introduction}

Pumping of the ground state main line masers at 1667 and 1665 MHz requires
asymmetry in the excitation of the two halves of the lambda doublets. The main
ingredients of the pump analysis are as follows:
\begin{enumerate}

\item Most molecules are in the ground state, from which they are excited into
higher levels and cascade back to the ground.

\item Downward transitions are due to radiative decays. For each level on the
$^2\Pi_{1/2}$ ladder, the strongest decay is always down the ladder. Thus
the cascade proceeds primarily along the $^2\Pi_{1/2}$ ladder, until the
final decay \PI(1/2,1/2) $\to$ \PI(3/2,3/2). Pumps that rely on
cross-ladder transitions are thus inherently weak.

\item
Line overlap modifies the fundamental radiative transition rates, making
the FIR pump cycle more conducive to asymmetries.

\end{enumerate}

The last point is especially important. In the absence of overlap, the escape
probability approximation for the radiative rate is $A \beta $, where $A$ and
$\beta$ are, respectively, the transition $A$-coefficient and the escape
probability. For optically thick transitions, $A \beta \simeq A/\tau$, and
the radiative decay rates then become independent of line strength. The
distinctions between different transitions disappear in this case, making it
difficult to maintain any asymmetry required for inversion. This is no longer
true in the presence of line overlap. The transition rate now depends on the
strengths of all overlapping lines in a complex way that involves also the
degree of overlap. The distinctions between different transitions are maintained
over a much larger range of column densities, creating the possibility for
stronger maser effect.

\begin{figure}
\centering\leavevmode
 \includegraphics[width=\figsize,clip]{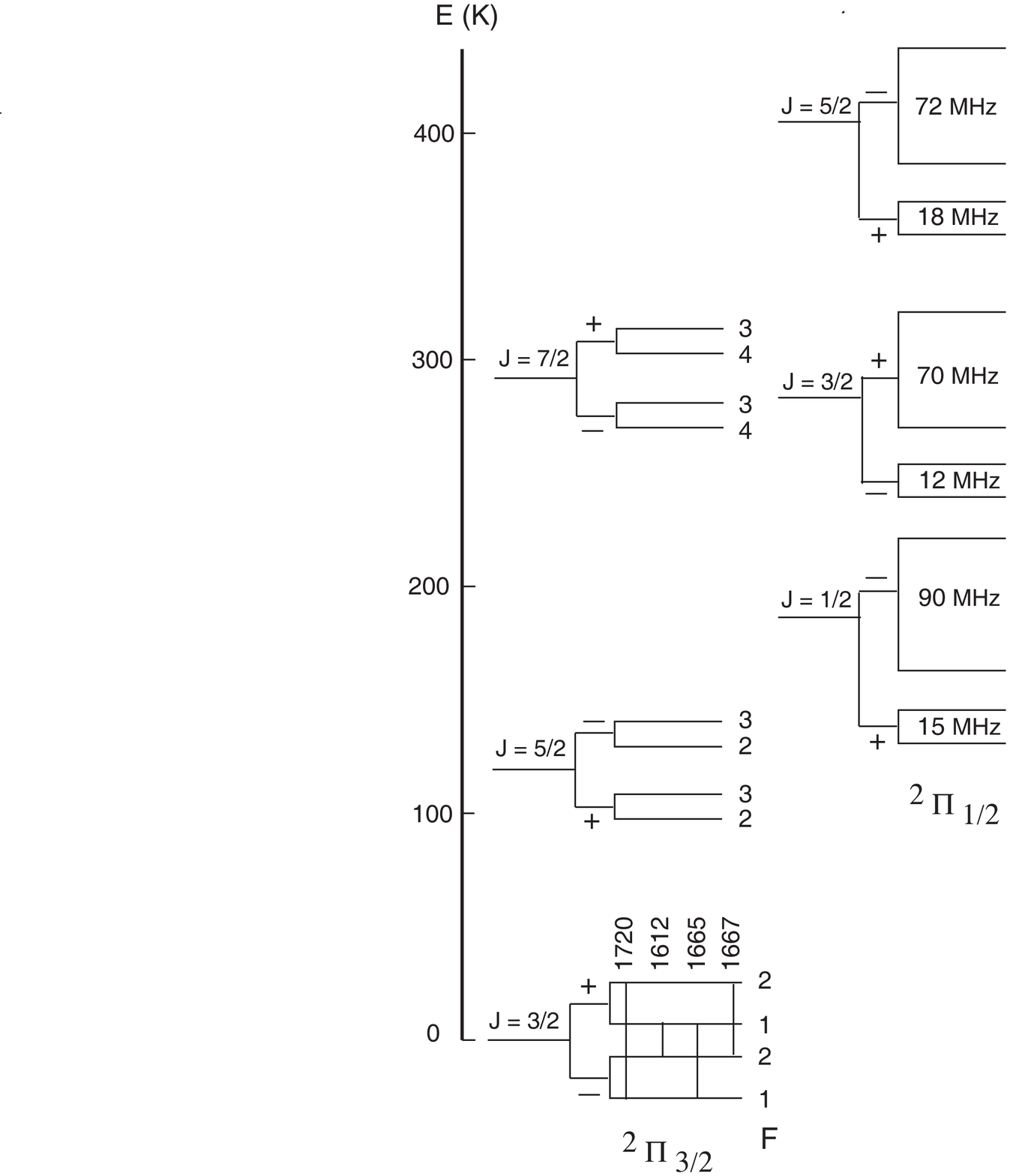}

\caption{The 24 hyperfine levels of OH used in our calculations.
$\Lambda$-doublet and hyperfine splitting are not to scale.  Note the very
uneven spacings in the lambda doublets of the $ ^2 \Pi _{1/2}$ ladder, which
cause extremely asymmetric line overlap between the FIR lines involving these
states.}
          \label{fig:levels}

\end{figure}

\begin{figure}
\centering\leavevmode
 \includegraphics[width=\figsize,clip]{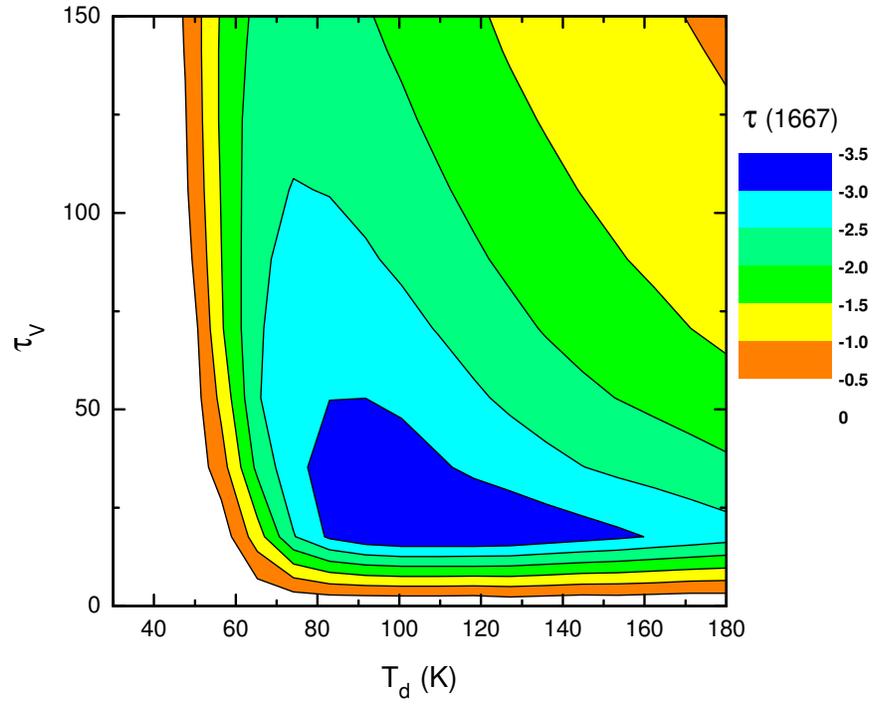}

\caption{Contour plots of the inverted 1667 MHz optical depth versus dust
temperature and optical depth. All other model parameters are as listed in
Table \ref{table:model}} \label{fig:tau contours}
\end{figure}

FIR main line pumping is dominated by line overlap effects, primarily in the 53 
\mic\ pump. On the $^2\Pi_{3/2}$ ladder, the hyperfine splits are roughly equal in
the two halves of the $\Lambda$-doublet. But on the $^2\Pi_{1/2}$ ladder, the 
hyperfine split of the upper half of each $\Lambda$-doublet is larger than in 
the lower half by factors of 4--6 (see figure \ref{fig:levels}). These uneven 
splits create an asymmetry in the overlap of the pump lines connecting the 
ground state to the rotational states of the $^2\Pi_{1/2}$ ladder. In contrast, 
there is essentially no overlap asymmetry in the image transitions for the 120 
\mic\ pump lines connecting the ground state to the the first excited state in 
the $^2\Pi_{3/2}$ ladder. The 53 \mic\ pump lines have special properties 
because they populate the $^2\Pi_{1/2}(J = 3/2)$ state, whose quantum numbers 
directly mirror the ground state. This places it in a special position during 
both the excitation and cascade phases. First, the 53 \mic\ pump cycle is the 
only one in which each upper level connects to the two ground state levels in 
the appropriate $\Lambda$-doublet; in all other cases, one of the excited 
rotational levels has only one transition to the ground because of the dipole 
selection rules. This doubles the number of potential overlaps compared to the 
other cycles, creating more asymmetry during the excitation phase of the pump 
cycle. Second, barring cross-ladder transitions, this is the only pump cycle 
that shuffles the molecules between the two halves of the ground-state 
$\Lambda$-doublet. In contrast, both the 80 \mic\ and the 35 \mic\ pump cycles 
return the molecules into the very same $\Lambda$-doublet from which they 
started.

The large efficiency of the 53 \mic\ pump is produced by asymmetric overlap
effects on both the absorption rates out of the ground state levels and the
returning decays. The inverting effects of line overlap on the absorption rates
were first noted by \citet{burdy1} and applied to OHMs by \cite{burdy2}. They
found that the smaller amount of overlap between the 53 \mic\ pump lines out of
the lower $\Lambda$-doublet of the ground state reduces the shielding from the
pumping dust radiation and thus increases the pumping out of the lower
$\Lambda$-doublet relative to that from the upper. \citet{burdy2} used a fixed
Doppler width of .7 \kms\ and found the 1667/1665 MHz flux ratio to be \about\
1, in agreement with our results.

The asymmetries of the radiative excitations disappear when \DV\ \ga\ 10 \kms\
because there is then complete overlap of the 53 \mic\ lines. However, the
inversion is sustained at larger linewidths due to overlap effects in the
cascade back to the ground. Overlaps among the \PI(1/2,3/2) $\to$ \PI(1/2,1/2)
cascades are particularly important since they are the primary decay routes out
of the upper rotational state of the 53 \mic\ pump. In addition, the overlap
asymmetry between the 2 halves of these $\Lambda$-doublets is extreme. The
separations between the transitions connecting the lower doublets range from .5
to 2.5 \kms, while the corresponding transitions between the upper doublets have
separations ranging from 3.3 to 15 \kms, a factor of 6 larger. The line overlap
asymmetry in these transitions remains important for very large linewidths. The
overlap of these lines significantly enhances the 53 \mic\ pump efficiency and
is necessary to invert the ground state $\Lambda$-doublets for linewidths \ga\
10 \kms.

When the linewidths exceed \about\ 40 \kms\ the asymmetries in the cascade
transitions disappear, too, and the ground state lines thermalize. It may be
noted that the calculations of \cite{burdy2} did not produce inversions at \DV
\ga\ 5 \kms\ because they neglected overlap effects in the decays. In contrast,
we include all possible overlaps, which essentially involve {\em every} FIR
transition when \DV\ \ga\ 10 \kms.

\clearpage

 \begin{figure}
\centering\leavevmode
\includegraphics[width=\figsize,clip]{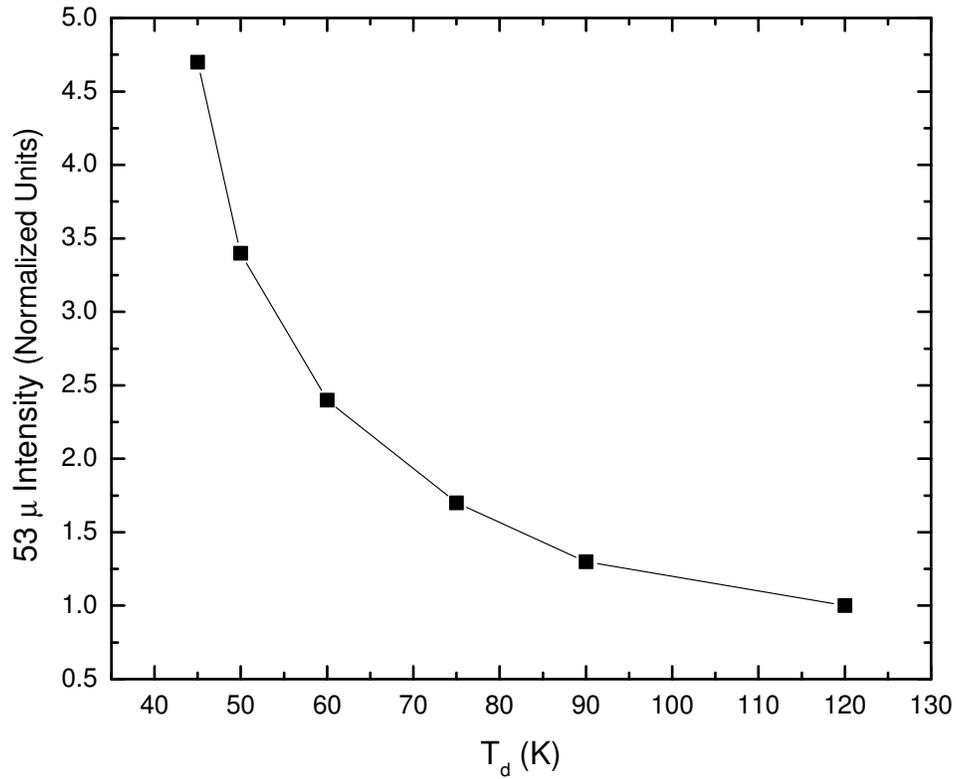}

\caption{Minimal 53 \mic\ intensity required for inversion --- cooler dust
requires larger minimal flux. At a given dust temperature, the plotted
intensity corresponds to the minimal $\tau_V$ that produces maser inversion
(see eq.\ 1). Results are normalized to the minimal intensity when $T_d = 120$
K (1.08 \x \E{-12} erg\,cm$^{-2}$\,s$^{-1}$\,Hz$^{-1}$\,ster$^{-1}$). Model
parameters are in Table \ref{table:model}.} \label{fig:minimal_F}

\end{figure}

\begin{figure}
\centering\leavevmode\includegraphics[width= \figsize,clip]{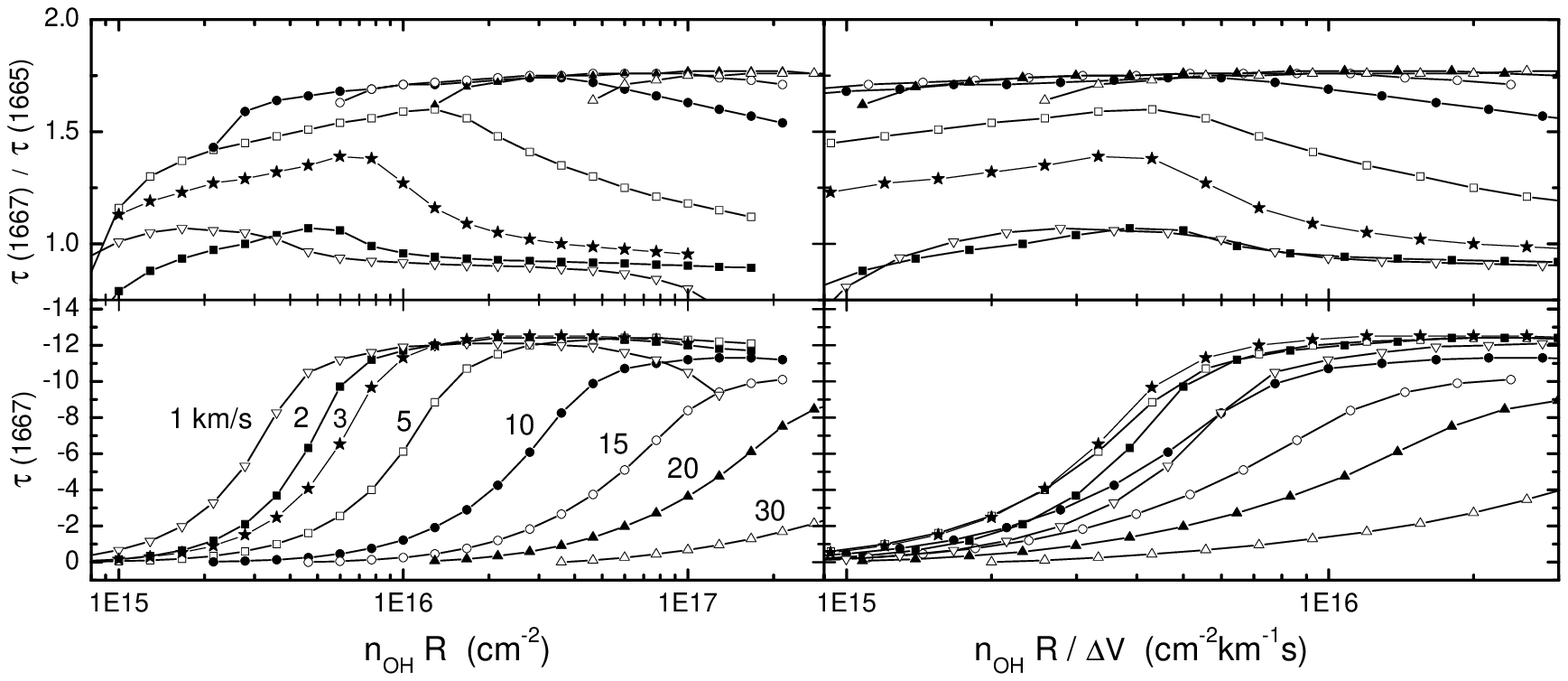}

\caption{1667 MHz optical depth and 1667/1665 MHz optical depth ratio for
different linewidths. Left panel is plotted versus OH column density. Right
panel is plotted versus OH column density per velocity interval. All other
model parameters from Table \ref{table:model}. Note that linewidths \la\ 2 
\kms\ lead to 1667/1665 MHz optical depth ratios \la\ 1, while linewidths \ga\ 3
\kms\ cause this ratio to increase dramatically, becoming \about 1.75 for
linewidths \ga\ 10 \kms.}
\label{fig:linewidth}

\end{figure}

\begin{figure}
\centering\leavevmode
\includegraphics[width=\figsize,clip]{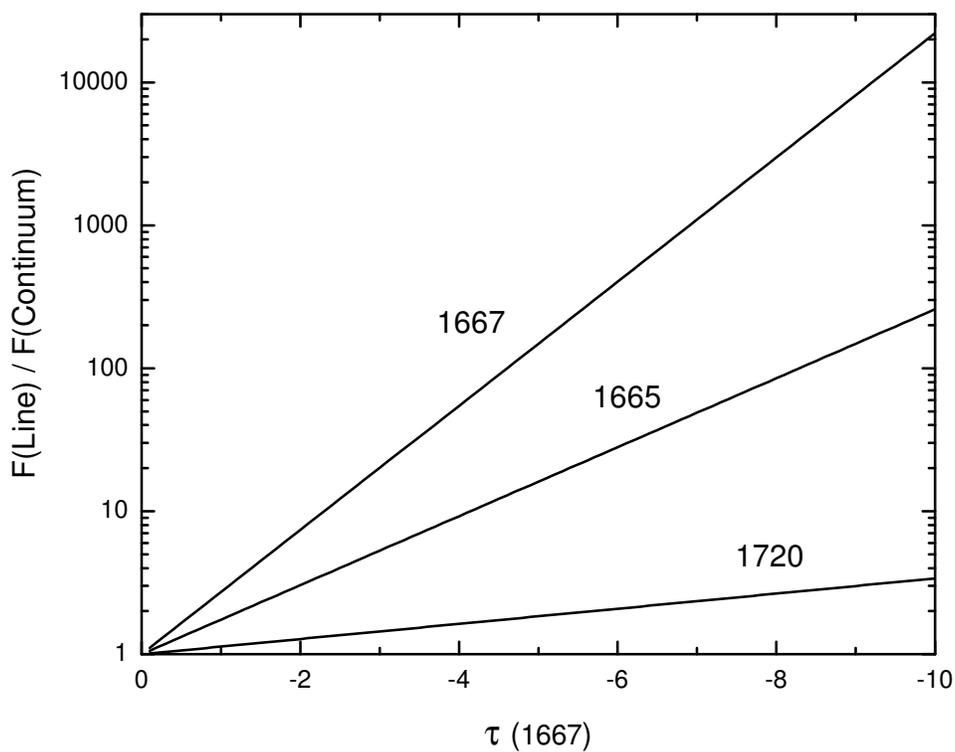}

\caption{Variation of flux with 1667 MHz optical depth for the three indicated masers
when all ground-state transitions have the same excitation temperature 
(the situation when $\Delta v$ \ga\ 10 \kms; see \S\ref{sec:overlap}). The 
1667/1665 and 1667/1720 flux ratios increase as the 1667 MHz maser gets stronger. 
The typically observed 1667/1665 MHz flux ratio of 5 implies $\tau(1667)\approx
-3.5$ and the 1720 MHz line will always be weak compared to the 1667 MHz line.}

\label{fig:fluxes}

\end{figure}

\begin{figure}
\centering\leavevmode \includegraphics[width=\figsize,clip]{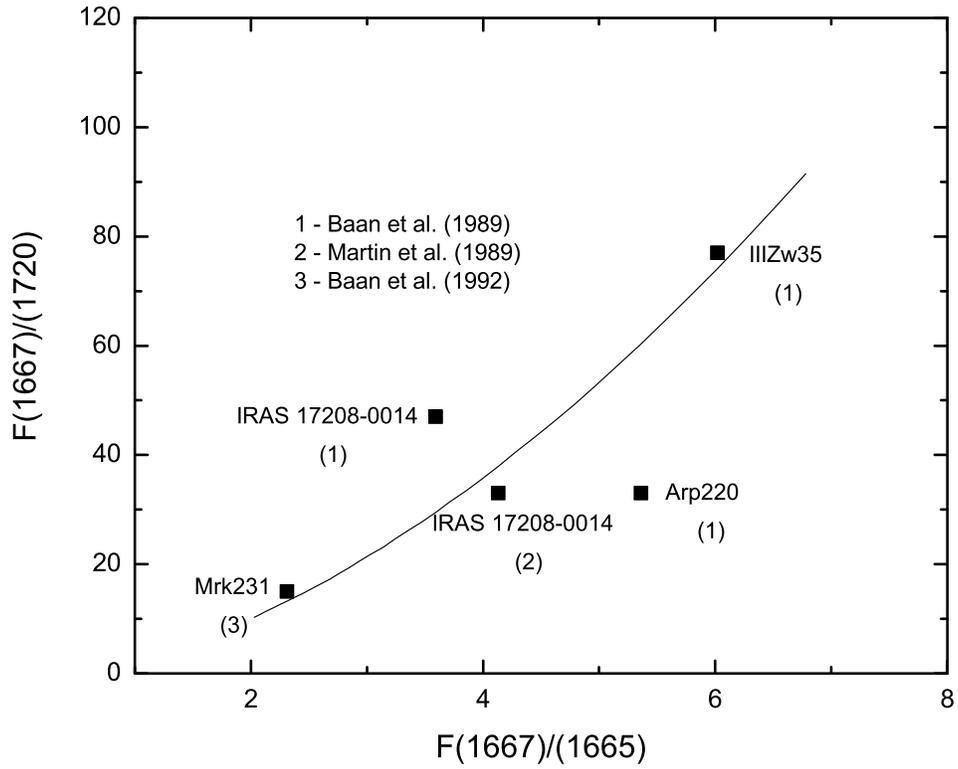}

\caption{``Color--color" diagram for OH maser lines. Filled squares show data
points for four OHMs. The solid line is the track of flux ratios of the
amplified background continuum when all lines have the same excitation
temperature. The 1667 MHz optical depth varies along the track from $-0.5$ at
the lower left to $-4$ at the upper right.} \label{fig:color-color}

\end{figure}

\begin{figure}
\centering\leavevmode \includegraphics[width=\figsize,clip]{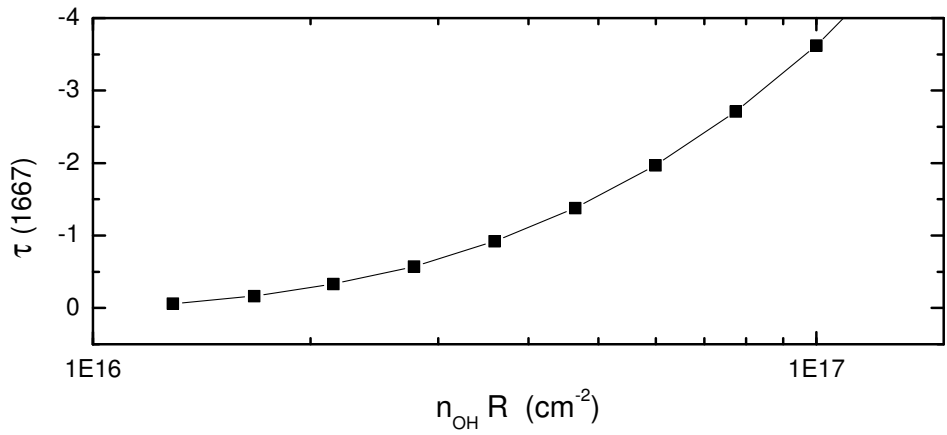}

\caption{1667 MHz maser optical depth for an OHM cloud using the
model parameters from Table \ref{table:model}. This is a zoom in on 
the relevant portion of figure 4. The optical depth of $-1.5$ for 
OH column density of \about\ 5\x\E{16}\cmsq\
is in good agreement with the requirements of the clumpy model of OHM 
emission in IIIZw35 (see \S\ref{sec:clumps}). }
\label{fig:clump}

\end{figure}


\begin{thebibliography}{}

\bibitem[Baan(1985)]{baan85} Baan, W.~A.\ 1985, \nat, 315, 26

\bibitem[Baan(1989)]{baan89} Baan, W.~A.\ 1989, \apj, 338, 804

\bibitem[Baan \& Haschick(1987)]{baan87} Baan, W.~A., \&
Haschick, A.~D.\ 1987, \apj, 318, 139

\bibitem[Baan et al.(1989)]{baanetal89} Baan, W.~A., Haschick,
A.~D., \& Henkel, C.\ 1989, \apj, 346, 680

\bibitem[Baan et al.(1992)]{baanetal92} Baan, W.~A., Haschick, A.,
\& Henkel, C.\ 1992, \aj, 103, 728

\bibitem[Blain et al.(2003)]{blain} Blain, A.~W., Barnard,
V.~E., \& Chapman, S.~C.\ 2003, \mnras, 338, 733

\bibitem[Burdyuzha \& Varshalovich(1973)]{burdy1} Burdyuzha,
V.~V., \& Varshalovich, D.~A.\ 1973, Soviet Astronomy, 17, 308

\bibitem[Burdyuzha \& Vikulov(1990)]{burdy2} Burdyuzha, V.~V.,
\& Vikulov, K.~A.\ 1990, \mnras, 244, 86

\bibitem[Capriotti(1965)]{capriotti} Capriotti, E.~R.\ 1965,
\apj, 142, 1101

\bibitem[Cernicharo et al.(2006)]{cernicharo} Cernicharo, J.,
Pardo, J.~R., \& Weiss, A.\ 2006, \apjl, 646, L49

\bibitem[Cesaroni \& Walmsley(1991)]{cesaroni} Cesaroni, R., \&
Walmsley, C.~M.\ 1991, \aap, 241, 537

\bibitem[Cragg et al.(2005)]{cragg} Cragg, D.~M., Sobolev,
A.~M., \& Godfrey, P.~D.\ 2005, \mnras, 360, 533

\bibitem[Darling(2007)]{darling07} Darling, J.\ 2007, \apjl, 669, L9

\bibitem[Darling \& Giovanelli(2002)]{darling02} Darling, J., \&
Giovanelli, R.\ 2002, \aj, 124, 100

\bibitem[Darling et al.(2003)]{darlingmethanol} Darling, J., Goldsmith,
P., Li, D., \& Giovanelli, R.\ 2003, \aj, 125, 1177

\bibitem[Downes \& Solomon(1998)]{downes} Downes, D., \&
Solomon, P.~M.\ 1998, \apj, 507, 615

\bibitem[Draine \& Lee(1984)]{drainelee84} Draine, B.~T., \& Lee,
H.~M.\ 1984, \apj, 285, 89

\bibitem[Elitzur(1992)]{elitzur} Elitzur, M. 1992, {\it Astronomical
     Masers}, (Dordrecht: Kluwer Academic Publishers)

\bibitem[Elitzur \& Asensio Ramos(2006)]{elitzur06} Elitzur, M.,
\& Asensio Ramos, A.\ 2006, \mnras, 365, 779

\bibitem[Flower(1990)]{flower} Flower, D. 1990, {\it Molecular
     Collisions in the Interstellar Medium}, (Cambridge: Cambridge
     University Press)

\bibitem[Gao \& Solomon(2004)]{gao} Gao, Y., \& Solomon,
P.~M.\ 2004, \apj, 606, 271

\bibitem[Gaume \& Mutel(1987)]{gaume} Gaume, R.~A., \& Mutel,
R.~L.\ 1987, \apjs, 65, 193

\bibitem[Graci{\'a}-Carpio et al.(2006)]{carpio}
Graci{\'a}-Carpio, J., Garc{\'{\i}}a-Burillo, S., Planesas, P., \& Colina,
L.\ 2006, \apjl, 640, L135

\bibitem[Greve et al.(2006)]{greve} Greve, T.~R.,
Papadopoulos, P.~P., Gao, Y., \& Radford, S.~J.~E.\ 2006, ArXiv
Astrophysics e-prints, arXiv:astro-ph/0610378

\bibitem[Guilloteau et al.(1981)]{guilloteau} Guilloteau, S.,
Lucas, R., \& Omont, A.\ 1981, \aap, 97, 347

\bibitem[Hartquist et al.(1995)]{hartquistetal} Hartquist, T.~W.,
Menten, K.~M., Lepp, S., \& Dalgarno, A.\ 1995, \mnras, 272, 184

\bibitem[Hartquist \& Sternberg(1991)]{hartquiststernberg} Hartquist,
T.~W., \& Sternberg, A.\ 1991, \mnras, 248, 48

\bibitem[He \& Chen(2004)]{he} He, J.~H., \& Chen, P.~S.\
2004, New Astronomy, 9, 545

\bibitem[Henkel et al.(1986)]{henkel86} Henkel, C., Guesten, R.,
\& Batrla, W.\ 1986, \aap, 168, L13

\bibitem[Henkel et al.(1987)]{henkel87} Henkel, C., Guesten, R.,
\& Baan, W.~A.\ 1987, \aap, 185, 14

\bibitem[Henkel \& Wilson(1990)]{henkelwilson90} Henkel, C., \&
Wilson, T.~L.\ 1990, \aap, 229, 431

\bibitem[Hinz \& Rieke(2006)]{hinz} Hinz, J.~L., \& Rieke,
G.~H.\ 2006, \apj, 646, 872

\bibitem[Kl{\"o}ckner(2002)]{klockner} Kl{\"o}ckner, H.-R., 2002,
     Ph.D Thesis, Rijksuniversiteit, Groningen

\bibitem[Kl{\"o}ckner et al.(2003)]{klockneretal} Kl{\"o}ckner,
H.-R., Baan, W.~A., \& Garrett, M.~A.\ 2003, \nat, 421, 821

\bibitem[Lisenfeld et al.(2000)]{lisenfeld00} Lisenfeld, U., Isaak,
K.~G., \& Hills, R.\ 2000, \mnras, 312, 433

\bibitem[Lo(2005)]{lo05} Lo, K.~Y.\ 2005, \araa, 43, 625

\bibitem[Lockett \& Elitzur(1989)]{lockettelitzur89} Lockett, P., \&
Elitzur, M.\ 1989, \apj, 344, 525

\bibitem[Lockett et al.(1999)]{lockettetal99} Lockett, P., Gauthier,
E., \& Elitzur, M.\ 1999, \apj, 511, 235

\bibitem[Lonsdale(2002)]{lonsdale02} Lonsdale, C.~J.\ 2002, Cosmic
Masers: From Proto-Stars to Black Holes, 206, 413

\bibitem[Lonsdale et al.(1998)]{lonsdale98} Lonsdale, C.~J.,
Lonsdale, C.~J., Diamond, P.~J., \& Smith, H.~E.\ 1998, \apjl, 493, L13

\bibitem[Lonsdale et al.(2003)]{lonsdale03} Lonsdale, C.~J.,
Lonsdale, C.~J., Smith, H.~E., \& Diamond, P.~J.\ 2003, \apj, 592, 804

\bibitem[Lonsdale et al.(2006)]{lonsdale06} Lonsdale, C.~J.,
Diamond, P.~J., Thrall, H., Smith, H.~E., \& Lonsdale, C.~J.\ 2006, \apj,
647, 185

\bibitem[Martin et al.(1989)]{martinetal89} Martin, J.~M., Le
Squeren, A.~M., Bottinelli, L., Gouguenheim, L., \& Dennefeld, M.\ 1989,
\aap, 208, 39

\bibitem[Momjian et al.(2006)]{momjian} Momjian, E., Romney,
J.~D., Carilli, C.~L., \& Troland, T.~H.\ 2006, \apj, 653, 1172

\bibitem[Offer et al.(1994)]{offeretal94} Offer, A. R., van Hemert,
     M. C., van Dishoeck, E. F. 1994, \jcp, 100, 362

\bibitem[Parra et al.(2005)]{parra05} Parra, R., Conway, J.~E.,
Elitzur, M., \& Pihlstr{\"o}m, Y.~M.\ 2005, \aap, 443, 383

\bibitem[Parra et al.(2007)]{parra07} Parra, R., Conway, J.~E.,
Diamond, P.~J., Thrall, H., Lonsdale, C.~J., Lonsdale, C.~J., \& Smith,
H.~E.\ 2007, \apj, 659, 314

\bibitem[Pihlstr{\"o}m et al.(2001)]{pihl} Pihlstr{\"o}m,
Y.~M., Conway, J.~E., Booth, R.~S., Diamond, P.~J., \& Polatidis, A.~G.\
2001, \aap, 377, 413

\bibitem[Randell et al.(1995)]{randell} Randell, J., Field, D.,
Jones, K.~N., Yates, J.~A., \& Gray, M.~D.\ 1995, \aap, 300, 659

\bibitem[Richards et al.(2005)]{richards} Richards, A.~M.~S.,
Knapen, J.~H., Yates, J.~A., Cohen, R.~J., Collett, J.~L., Wright, M.~M.,
Gray, M.~D., \& Field, D.\ 2005, \mnras, 364, 353

\bibitem[Skinner et al.(1997)]{skinner} Skinner, C.~J., Smith,
H.~A., Sturm, E., Barlow, M.~J., Cohen, R.~J., \& Stacey, G.~J.\ 1997,
\nat, 386, 472

\bibitem[Soifer et al.(1999)]{soiferarp} Soifer, B.~T.,
Neugebauer, G., Matthews, K., Becklin, E.~E., Ressler, M., Werner, M.~W.,
Weinberger, A.~J., \& Egami, E.\ 1999, \apj, 513, 207

\bibitem[Thompson et al.(2005)]{thompson} Thompson, T.~A.,
Quataert, E., \& Murray, N.\ 2005, \apj, 630, 167


\bibitem[Wardle(1999)]{wardle} Wardle, M.\ 1999, \apjl, 525, L101

\bibitem[Yun \& Carilli(2002)]{yuncarilli} Yun, M.~S., \& Carilli,
C.~L.\ 2002, \apj, 568, 88


\end{thebibliography}
\end{document}